# Tunable spin transport using zigzag MoS$_2$ nanoribbon

Hamidreza Simchi,[1,2] Mahdi Esmaeilzadeh,[1] Hossein Mazidabadi,[1] Mohammad Norouzi,[1]

[1]Department of Physics, Iran University of Science and Technology, Narmak, Tehran 16844, Iran

[2]Semiconductor Technology Center, Tehran Iran

We consider a zigzag nanoribbon of MoS$_2$ and study its spin-dependent conductance and spin-polarization in presence (absent) of an external electric field. The field is not only perpendicular to the molybdenum plane (called vertical field $E_z$) but also to the transport direction of electrons (called transverse field $E_t$). It is shown that, while in the absent of the external electric field, two bands of seven bands are spin split in the valence band but no spin splitting is seen at point $k_x = 0.0$. Under applying the electric field we show, the ribbon behaves as a prefect spin up (down) filter if and only if the both components of the electrical field are present. Finally it is shown that, by changing the strength of $E_t$, the ribbon acts as a spin inverter for fixed values of $E_z$.

**Keywords:** MoS$_2$ nanoribbon, non-equilibrium Green's Function, Spin transport, Spin filter, Spin inverter

I. **Introduction**

Some bulk materials are composed by stacks of strongly layers with weak interlayer attraction, allowing exfoliation into individual, atomically thin layers.[1] Transition metal dichalcogenides (TMDCs) with formula MX$_2$, where M is a transition metal and X is a chalcogen, are examples of these kinds of bulk materials.[2-3] They are known as two dimensional (2D) materials which shows a wide range of electronic, optical, mechanical, chemical and thermal properties.[2-6] Molybdenum disulfide (MoS$_2$) is a TMDC semiconductor with an indirect band gap.[7] The MoS$_2$ monolayer is a semiconductor with the computed direct $K \to K$ gap equal to 1.79 eV and the computed indirect $\Gamma \to Q$ gap equal to 2.03 eV.[8] The computed band gap is in good agreement with the experimentally measured emission energy at 1.83-1.98 eV.[9,10] Cao et al., have shown that, monolayer of MoS$_2$ is an ideal material for valleytronics.[11] It has been shown that, inversion symmetry breaking together with spin-orbit coupling leads to coupled spin and valley degree of freedoms in monolayer of MoS$_2$ and other group-VI dichalcogenides.[12] It means that, it is possible to control both spin and valley in these 2D materials.[12] Peelaers et al., have shown that, strain allows engineering the magnitude as well as the nature (direct versus indirect) of the band gap in MoS$_2$.[13] Also, it has been shown that, with increasing the strain the exciton binding energy is nearly unchanged, while optical gap is reduced significantly in monolayer of MoS$_2$.[14] Kang et al., have studied the band offsets and the heterostructures of monolayer and few-layer of MX$_2$ and it has been shown that some of the monolayers and their heterostructures are suitable for potential application in optoelectronics.[15] One dimensional (1D) nanostructures have been of both fundamental and technological interest due to the interesting electronic and physical properties intrinsically associated with their low dimensionality and quantum confinement effect. Using spin-polarized first principle density functional theory (DFT) method, Li et al., have shown that zigzag (armchair) nanoribbons of MoS$_2$ show the ferromagnetic (nonmagnetic) and metallic (semiconductor) behavior.[16] Also, it has been shown that, armchair nanoribbons of MoS$_2$ could be metallic and exhibit a magnetic moment; and, when passivating with hydrogen, they become semiconductor.[17] Zigzag nanoribbons of MoS$_2$ are metallic and exhibit unusual magnetic properties regardless of passivation.[17] Sometimes, the band gap of material can be tuned by applying an external electric field. Using first-principle calculation, Yue et al., have shown that, the band gap of monolayer armchair MoS$_2$ nanoribbons can be significantly reduced and be closed by applying a transverse field, whereas the band gap modulation is absent under perpendicular field.[18] It has been shown that, tensile strain in the zigzag direction of single-layer MoS$_2$ zigzag nanoribbon produces reversible modulation of magnetic moments and electronic phase transition.[19] The strain-induced modulation can be enhanced or suppressed further by applying an electric field.[19]

In this paper, we consider a monolayer zigzag naoribbon of MoS$_2$ and study its spin-dependent conductance and spin-polarization in presence (absent) of an external electric field which is not only perpendicular to the molybdenum (Mo) plane (called vertical field $E_z$) but also to the transport direction of electrons (called transverse field $E_t$). It is shown that in the absent of the field, while two bands of seven bands structure are spin split in the valence band but no spin splitting is seen at point $k_x = 0.0$. We show that, the spin splitting is happened in the lowest unoccupied molecular orbital (LUMO) region of transmission curve under applying the field, and zigzag MoS$_2$ nanoribbon behaves as a prefect spin up(down) filter. Also it is shown that, the device acts as a prefect spin filter if and only if both components of the field are present. Finally, we show that, the ribbon behaves as a spin inverter by changing the transverse electric filed, $E_t$ for fixed value of vertical electric field, $E_z$. The organization of paper is as follow. In section II, the calculation method is presented. The results and discussion are presented in section III and conclusion and summary are provided in section IV.

II. **Calculation method**

Different tight binding Hamiltonians have been introduced for monolayer of MoS$_2$ before.[20-24] When the external electric field is present, there is no symmetry between two sulfur (S) atoms and the relevant orbitals accounting for the band structure are the $4d-orbitals$ of the Mo, $d_{x^2-y^2}, d_{xy}$ and $d_{z^2}$, and two $3p-orbitals$ of S, $p_x, p_y$.[22] The $7 \times 7$ Hamiltonian ($H$) and overlap ($S$) matrices of MoS$_2$ is written as:[22]

$$H(k) = \begin{pmatrix} H_a & H_t & H_t \\ H_t^\dagger & H_b & 0 \\ H_t^\dagger & 0 & H_{b'} \end{pmatrix}, \quad S = \begin{pmatrix} 1 & S & S \\ S^\dagger & 1 & 0 \\ S^\dagger & 0 & 1 \end{pmatrix} \quad (1)$$

with the on-site energy Hamiltonian[22], $H_a$, $H_b$, $H_{b'}$ and hopping matrix[22]

$$H_t = \begin{pmatrix} t_{11} f(k,\omega) & -e^{-i\omega} t_{11} f(k,-\omega) \\ t_{21} f(k,-\omega) & t_{22} f(k,0) \\ -t_{22} f(k,0) & -e^{-i\omega} t_{21} f(k,\omega) \end{pmatrix} \quad (2)$$

where $t_{11}$ is hopping integral between orbital $d_{z^2}$ of Mo-atom and orbital $p_x+ip_y, p_x-ip_y$ of S-atom. Also, $t_{21}$ is hopping integral between orbital $d_{x^2-y^2}+id_{xy}$ ($d_{x^2-y^2}-id_{xy}$) of Mo-atom and orbital $p_x+ip_y$ ($p_x-ip_y$) of S-atom. Finally, $t_{22}$ is hopping integral between orbital $d_{x^2-y^2}+id_{xy}$ ($d_{x^2-y^2}-id_{xy}$) of Mo-atom and orbital $p_x-ip_y$ ($p_x+ip_y$) of S-atom.

$f(\vec{k},\omega) = e^{i\vec{k}\cdot\vec{\delta}_1} + e^{i(\vec{k}\cdot\vec{\delta}_2+\omega)} + e^{i(\vec{k}\cdot\vec{\delta}_3-\omega)}$ is the structure factor with $\omega = 2\pi/3$, in plane momentum $\vec{k}=(k_x, k_y)$, and in-plane components of the lattice vector $\vec{\delta}_{i\pm}$ ($i=1,2,3$).[22] The overlap matrix $S$ id defined similar to $H_t$ but with hopping integral $s_{\mu\nu} = 0.1 t_{\mu\nu}$.[22] The spin-orbit coupling term (SOCT) $H_{SO}^{Mo} = \lambda\, diag(0, s, -s)$, where $\lambda = 0.08$ eV is the spin-orbit coupling constant and $s=\pm 1$.[22] It should be noted that the most important contribution of Mo atoms is considered in SOCT.[22] The values of on-site energies and hopping integrals and also vectors $\vec{\delta}_{i\pm}$ have been given in Ref.22 and thus we do calculate them in the present study.

For calculating the spin-dependent conductance and spin polarization, we consider a linear chain of unit cells (in x-direction), including 8 Mo-atoms and 16 S-atoms such that, the unit cells of left lead (L) are placed at positions $-\infty$.....-3, -2, -1, 0, the transport channel (C) is placed at positions 1, 2, .....,n-1,n, and the unit cells of right lead (R) are placed at positions n+1, n+2,n+3,....... $+\infty$ (see Fig.1). The wide of ribbon is in the y-direction and z-direction is perpendicular to the Mo-plane. The transport channel is composed by ten unit cells and includes 240 atoms totally. The transverse electric field ($E_t$) creates a potential difference between atoms with different y-value. The vertical electric field ($E_z$) creates a potential difference between up and down sulfur atoms. The both potential differences are added to the on-site energy of atoms. Of course it is assumed that, the external electric field acts only on the channel region and does not act on the left and right leads. The surface Green's function of left ($g_L$) and right ($g_R$) leads are calculated by suing the Sancho's algorithm.[25] The self energies, $\sum_{L(R)}$, and the coupling matrices, $\Gamma_{L(R)}$, can be calculated by using the surface Green's function.[26-28] The Green's function of the ribbon can be calculated by using[26]

$$\xi^{\uparrow(\downarrow)} = ((E+i\eta) \times I - H_C^{\uparrow(\downarrow)} - \sum_L - \sum_R)^{-1} \tag{3}$$

where $E$ is electron energy, $\eta$ is an infinitesimal number, $I$ is the unit matrix, and $\uparrow(\downarrow)$ stands for spin up (down).[26-28]

Now, the spin dependent conductance can be calculated by using[26-28]

$$\begin{aligned}G^{\uparrow} &= \text{Im}(Trace(\Gamma_L \times \xi^{\uparrow} \times \Gamma_R \times \xi^{\uparrow\dagger})), \\ G^{\downarrow} &= \text{Im}(Trace(\Gamma_L \times \xi^{\downarrow} \times \Gamma_R \times \xi^{\downarrow\dagger})),\end{aligned} \tag{4}$$

where, Im denotes the imaginary part. The spin-polarization is defined by[26,27,28]

$$P_S \equiv \frac{(G^{\uparrow} - G^{\downarrow})}{(G^{\uparrow} + G^{\downarrow})} \tag{5}$$

III. **Results and discussion**

Fig. 1 shows a monolayer zigzag nanoribbon of $MoS_2$. The channel includes 80 Mo-atoms and 160 S-atoms (totally 240 atoms). Mo-atoms and S-atoms are shown in blue and yellow color, respectively. Fig.2 (a) shows the seven band structure of monolayer of $MoS_2$ for both spins up and down which is found by using Eq. (1) and Eq. (2) when the electric field is absent. It is in good agreement with the five band structure of Ref.22. As the figure shows, the spin splitting is seen in the valence band. Of course at point $k_x = 0.0$, no spin splitting is seen in the valence band and energy band gap $E_g = 5.338$ eV. It should be noted that, $a_0 = a\cos(\theta)$ where, $a = 2.43 A^0$, and $\theta = 40.7$ degree. $a$ is the Mo-S bond length and $\theta$ is the angle between the bond and the Mo's plane.[22] Fig. 2(b) shows the conductance versus electron energy of the zigzag MoS2 nanoribbon at point $k_x = 0.0$ when the external electric field is absent. As it shows, no spin splitting is seen and transmission gap is equal to 5.6 eV which is at order of $E_g$ at this point.

Fig.3 shows the conductance versus electron energy when $E_t = -2$ V/nm and $E_z = 2$ V/nm. As it shows, spin splitting is happened in the lowest unoccupied molecular orbital (LUMO) region. It is happened since transmission gap is closed by electric fields at least for one of two components of spin for some values of electron energy. For example, when $E = 1.48$ eV then $G^\uparrow = 0.0(G_0)$ and $G^\downarrow = 1.48(G_0)$ while when $E = 1.7$ eV then $G^\uparrow = 3.013(G_0)$ and $G^\downarrow = 0.0(G_0)$. Fig.4 shows spin polarization versus electron energy when $E_t = -2$ V/nm and $E_z = 2$ V/nm. As it shows, the device can habit as a prefect spin down (up) filter i.e., $P_S = -1(+1)$ when $E = 1.48(1.7)$ eV. It is happened due to the gap closing by applying the electric field.

Fig.5 (a) shows the three dimensional graph of spin polarization versus electron energy and transverse field $E_t$ when $E_z = 2$ V/nm. Also, Figs. 5(b) and (c) show the conductance of spin up, $G^{\uparrow}$ and spin down $G^{\downarrow}$ versus electron energy and transverse field $E_t$ when $E_z = 2$ V/nm. By attention to these figures, it can be concluded that $P_S = \pm 1$ for a wide range of $E_t$ and electron energy when $E_z = 2$ V/nm. Also as Fig.5 (a) shows, when $E_t \geq -1.5$ V/nm $P_S = 0.0$ when $E = 1.48 (1.7)$ eV.

Lets us to focus on the two electron energies $E = 1.48$ eV and $E = 1.7$ eV and study the effect of vertical and transverse electric fields on the spin polarization at these points. We assume that, not only $E_t$ but also $E_z$ changes and study the effect of their variations on the spin polarization when $E = 1.48$ eV and $E = 1.7$ eV. Fig.6 shows spin polarization versus transverse electric field $E_t$ for different values of perpendicular electric field $E_z = -2, -1, 0.0, 1, 2$ V/nm when electron energy $E = 1.48$ eV. As it shows, when $E_z = 0.0$ V/nm no spin polarization is seen for all values of $E_t$ (solid line in green color). Therefore, both electric fields are necessary for creating the spin polarization. Also as it is seen, when $E_z = +2 \quad and \quad E_t = -2$ V/nm and $E_z = -2 \quad and \quad E_t = -1.95$ V/nm, the spin polarization $P_S = -1$, approximately. But, when $E_z = +2 \quad and \quad E_t = -1.72$ V/nm and $E_z = -2 \quad and \quad E_t = -1.76$ V/nm, the spin polarization $P_S = +1$, approximately. Therefore, by changing $E_t$ and fixing $E_z$ one can invert spin polarization perfectly. Fig.7 is similar to Fig.6 but here, the electron energy $E = 1.7$ eV. Although, both electric field s are necessary for creating the spin polarization, but here, the prefect spin inversion is not seen.

Fig.8(a) shows three dimensional graph of spin polarization and Figs.8 (a) and (b) show three dimensional graph of spin up conductance $G^\uparrow$ and spin down conductance $G^\downarrow$ versus transverse and perpendicular electric fields when electron energy $E = 1.48$ eV, respectively. As Figs.8 (b) and (c) show, only for some restricted values of $E_t$ one of the two components of spin up and down conductance is not zero and therefore only for these values of conductance the spin polarization cannot be ignored. Fig.9 (a) is similar to the Fig. 8 but here, the electron energy $E = 1.7$ eV. Similar to $E = 1.48$ eV, for some restricted values of $E_t$ one of the two components of spin up and down conductance is not zero and therefore only for these values of conductance the spin polarization cannot be ignored.

IV.     Conclusion and summary

We have considered a monolayer zigzag ribbon of MoS$_2$ and studied its spin-dependent conductance and polarization by using the tight binding non-equilibrium Green's function method in presence (absent) of and external electric filed. The field was not only perpendicular to the Mo-plane (vertical filed) but also to the transport direction of electrons (transverse field). In absence of the electric filed, it has been shown that, while the two bands of seven bands structure are spin split in the valence band but no spin splitting is seen in transmission curve at point $k_x = 0.0$. We have shown that, under applying the electric field, the MoS$_2$ ribbon behaves as a prefect spin up (down) filter duo to the transmission gap closing by the electric field. Also, it has been shown that, the device can habits as a prefect spin filter if and only if both transverse and vertical electric field is present. Finally, we have shown that under fixed value of vertical electric field the MoS$_2$ nanoribbon could act as spin inverter by changing the strength of the transverse electric field.

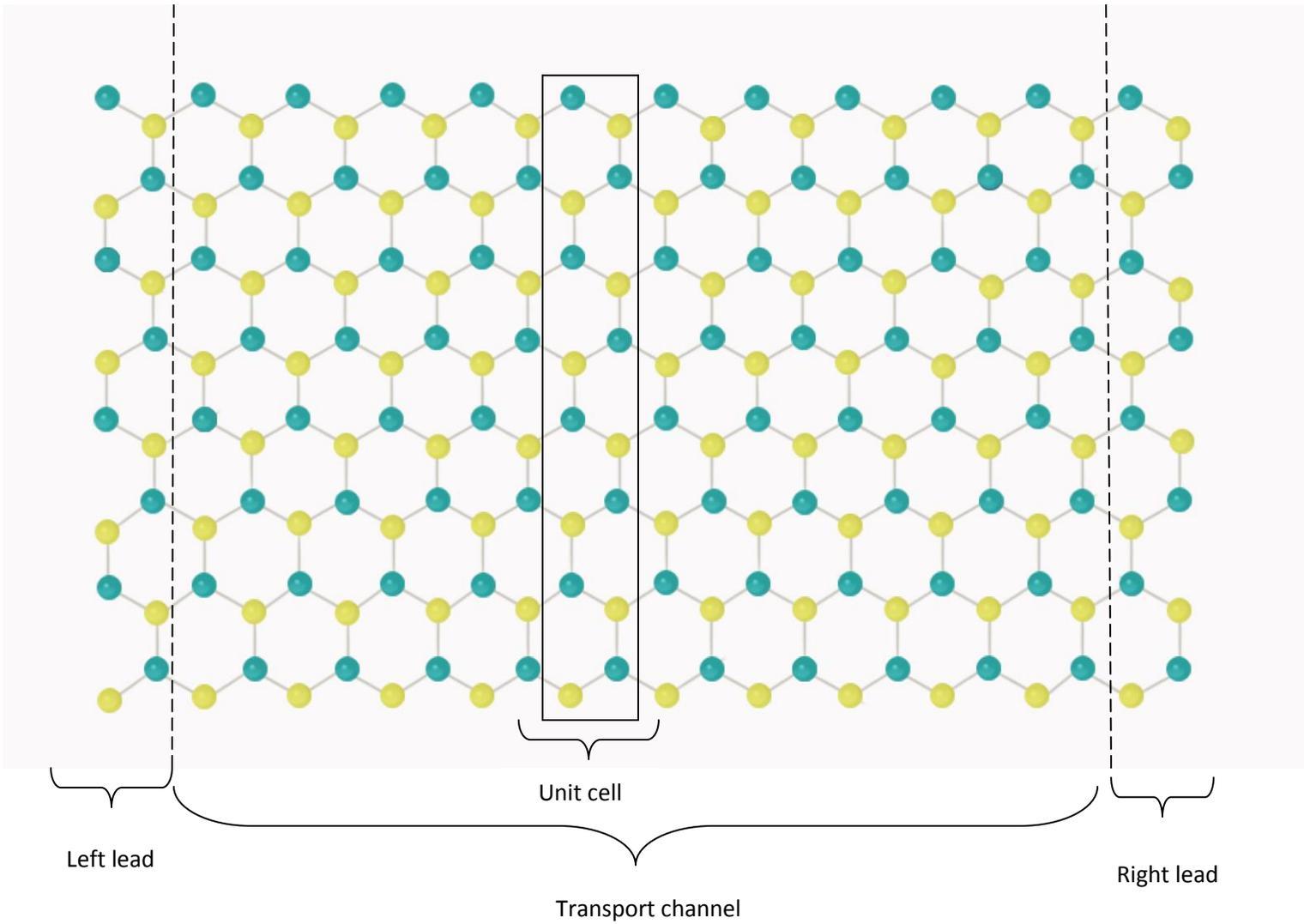

Fig. 1 A zigzag monolayer of MoS$_2$ nanoribbon. Each unit cells includes 8 Mo-atoms and 16 S-atoms and therefore, transport channel is composed by 240 atoms. The channel is connected to semi-infinite monolayer nano-ribbons of MoS$_2$ as left and right leads. Mo-atoms and S-atoms are shown in green and yellow color, respectively.

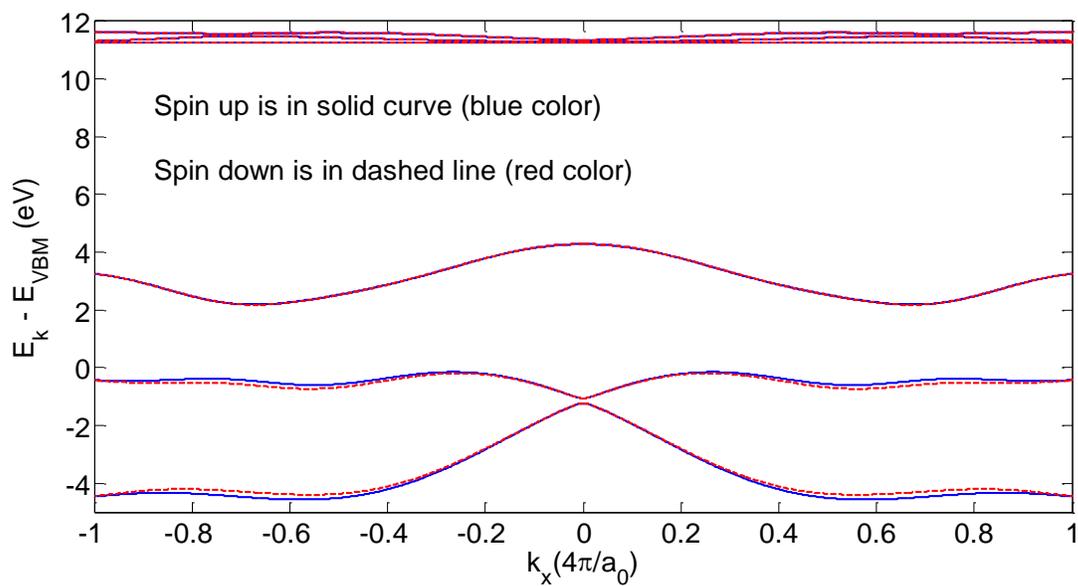

(a)

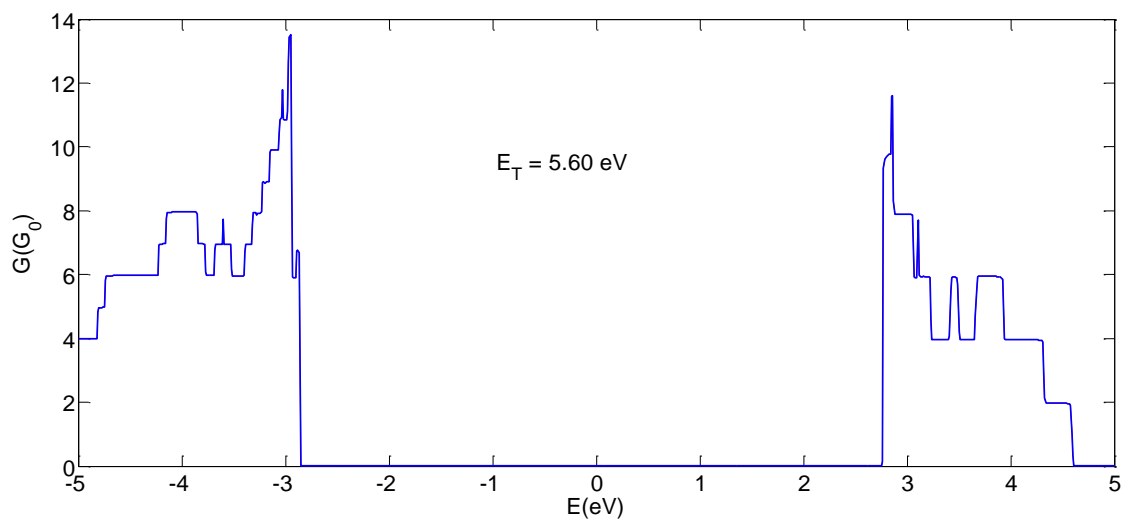

(b)

Fig. 2 (Color online) (a) Band structure of monolayer of MoS$_2$ consisting of seven bands in which two are spin split in the valence band. (b) Conductance versus electron energy of monolayer of zigzag nanoribbon of MoS$_2$ at $k_x = 0.0$ (the number of unit cells in each super cell is $N = 8$). $a_0 = a\cos(\theta)$ where, $a = 2.43\,\text{A}^0$, and $\theta = 40.7$ degree. $a$ is the length of Mo-S bond and $\theta$ is the angle between the bond and the $xy$ plane.[22]

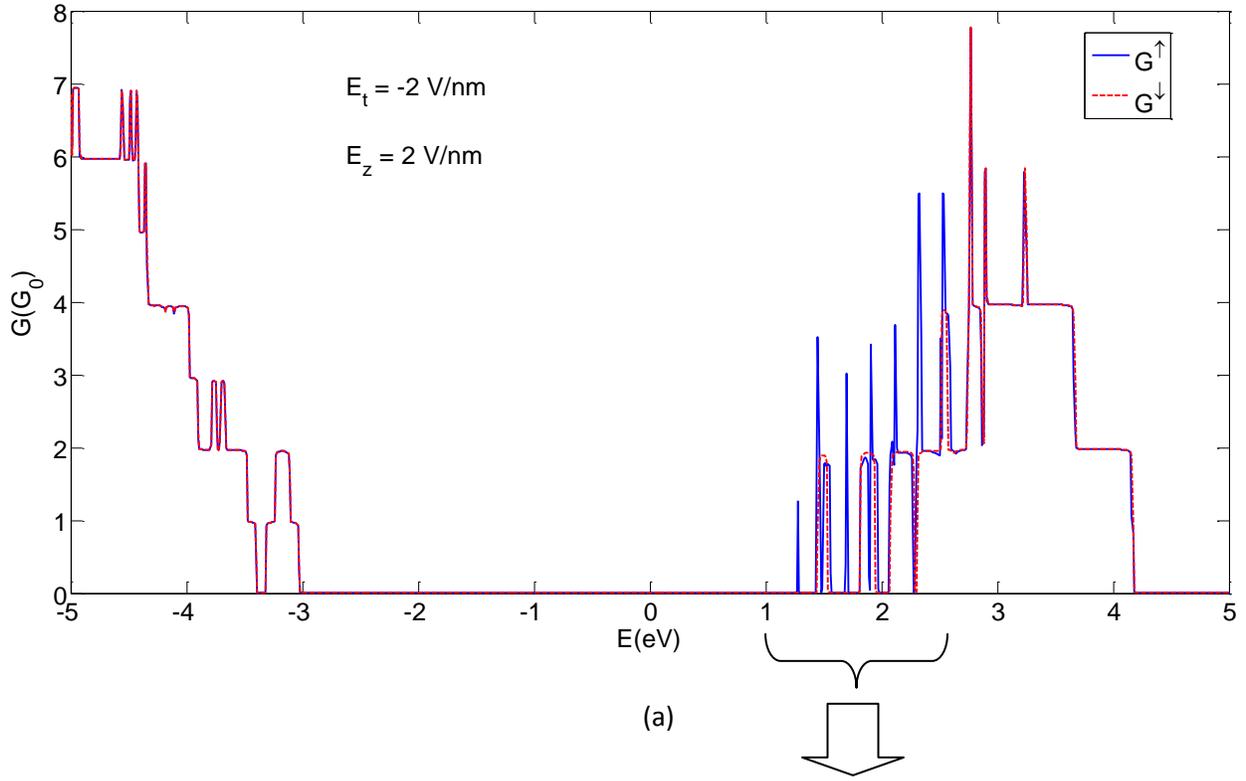

(a)

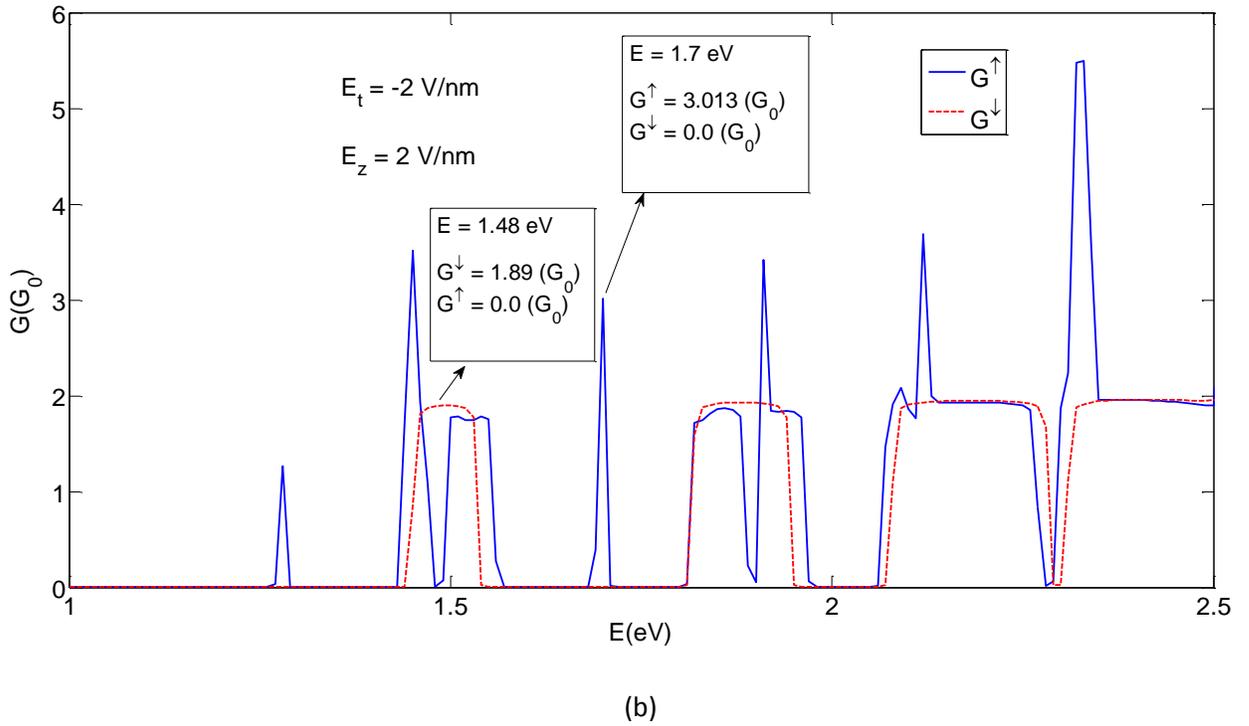

(b)

Fig. 3 (Color online) Conductance versus electron energy of monolayer zigzag anaoribbon of $MoS_2$ when perpendicular electric field $E_z$ and transverse electric field $E_t$ are present.

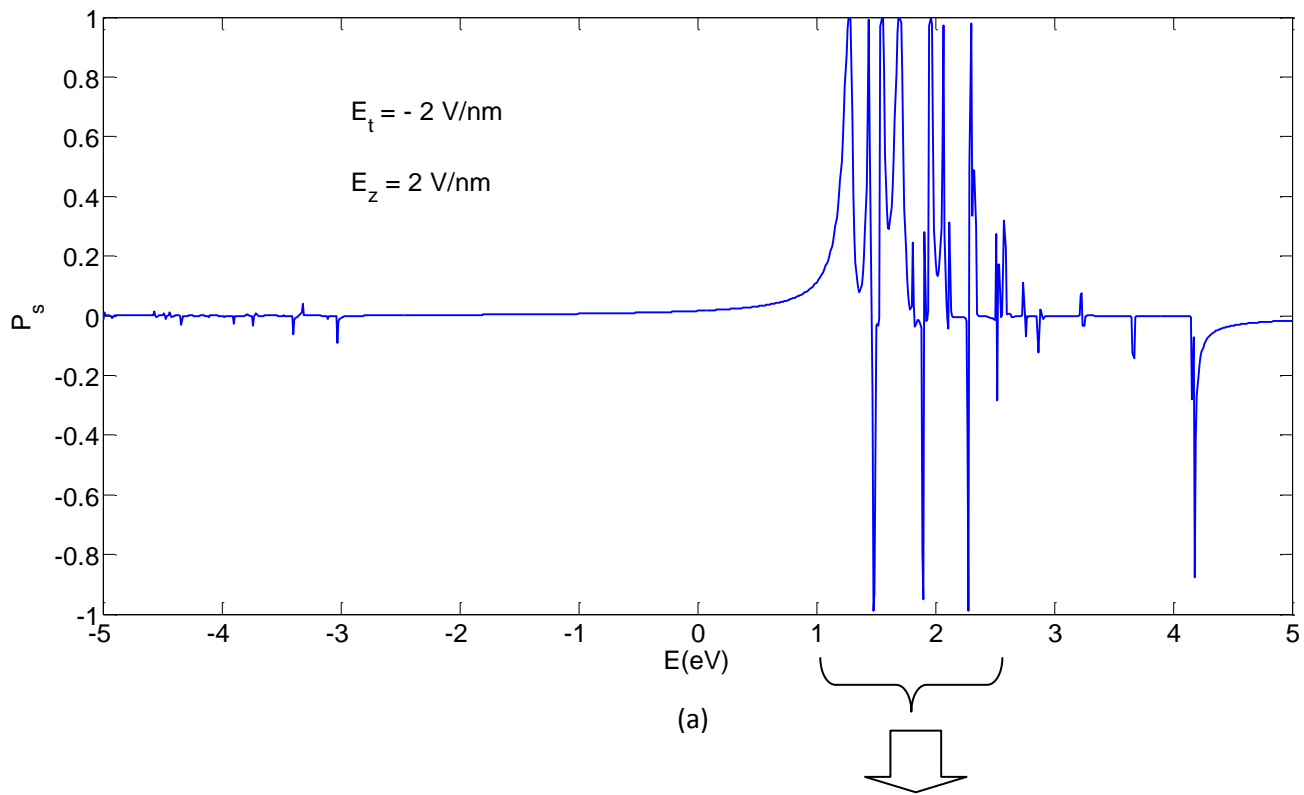

(a)

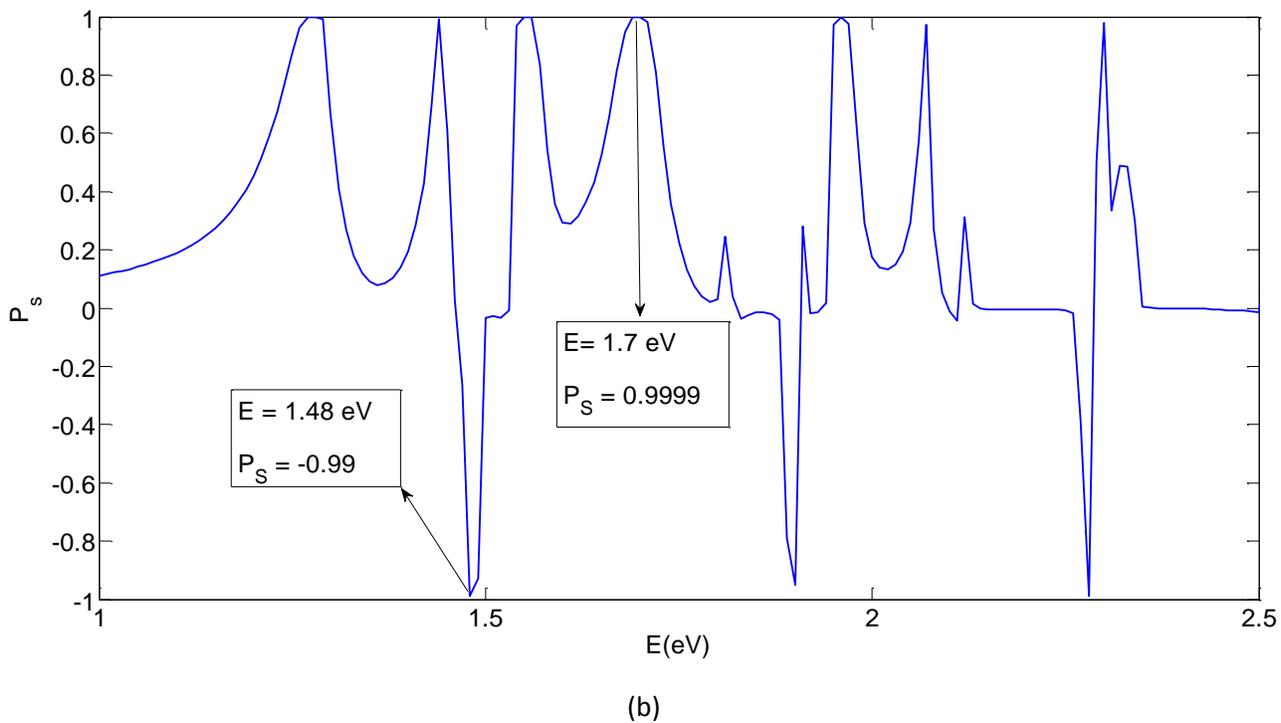

(b)

Fig. 4 (Color online) Spin polarization versus electron energy of monolayer zigzag anaoribbon of MoS$_2$ when perpendicular electric field $E_z$ and transverse electric field $E_t$ are present.

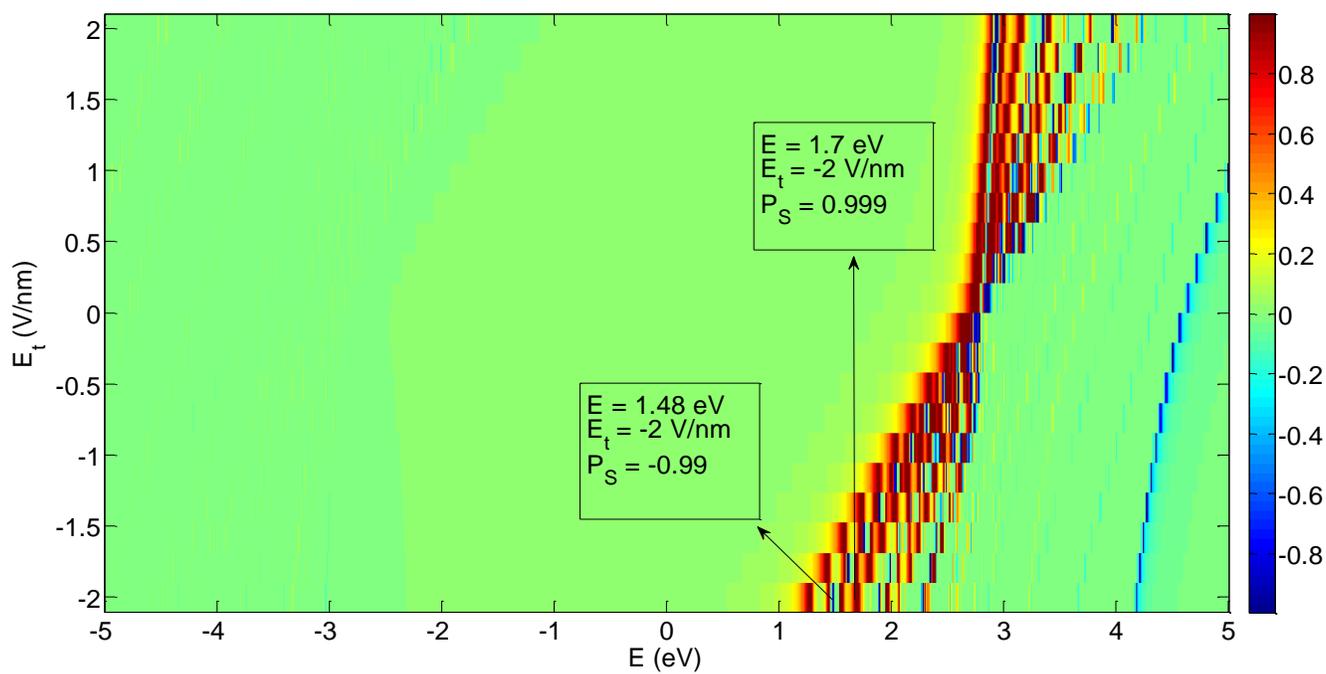

(a)

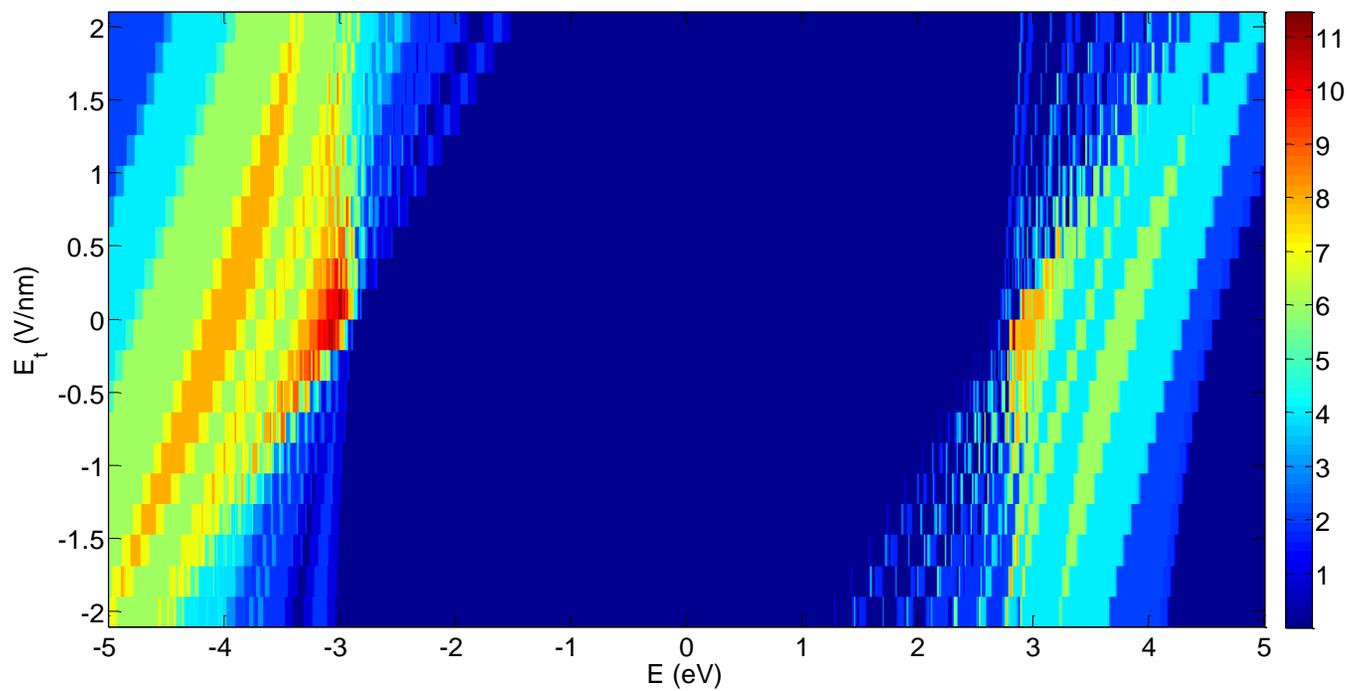

(b)

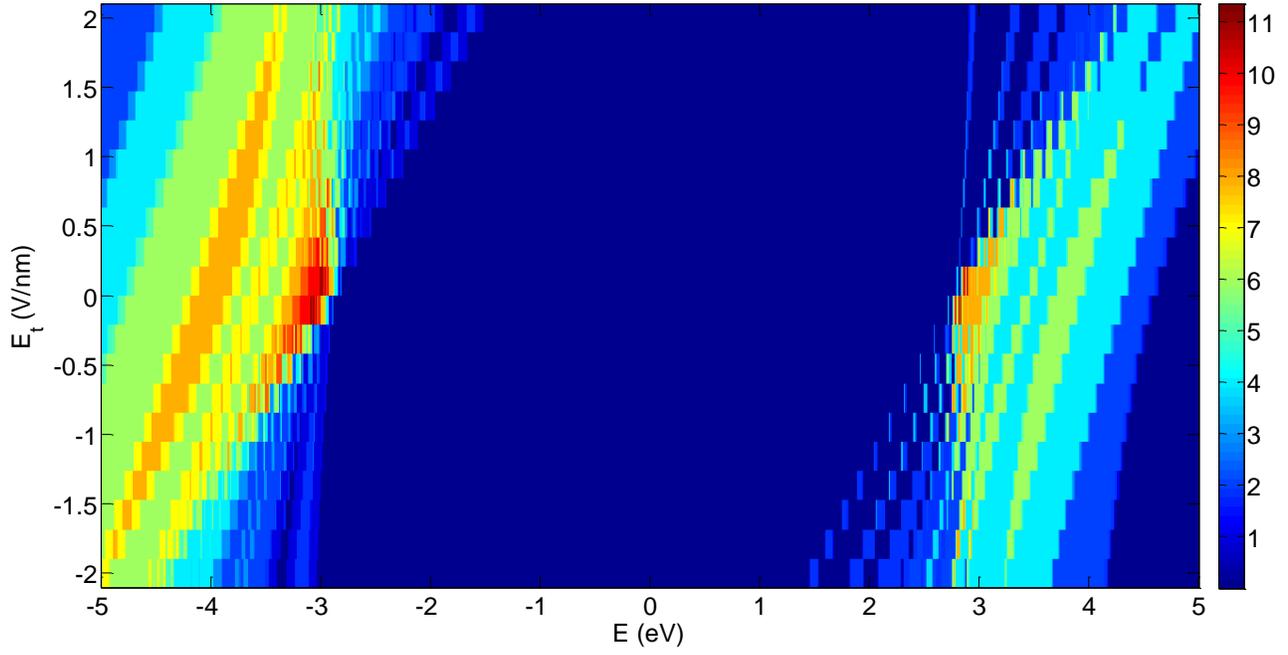

Fig. 5 (Color online) (a) Spin polarization, (b) conductance of spin up, $G^{\uparrow}$ and (c) conductance of spin down, $G^{\downarrow}$ electrons versus electron energy and transverse electric field, $E_t$ when perpendicular electric field $E_z = 2$ V/nm.

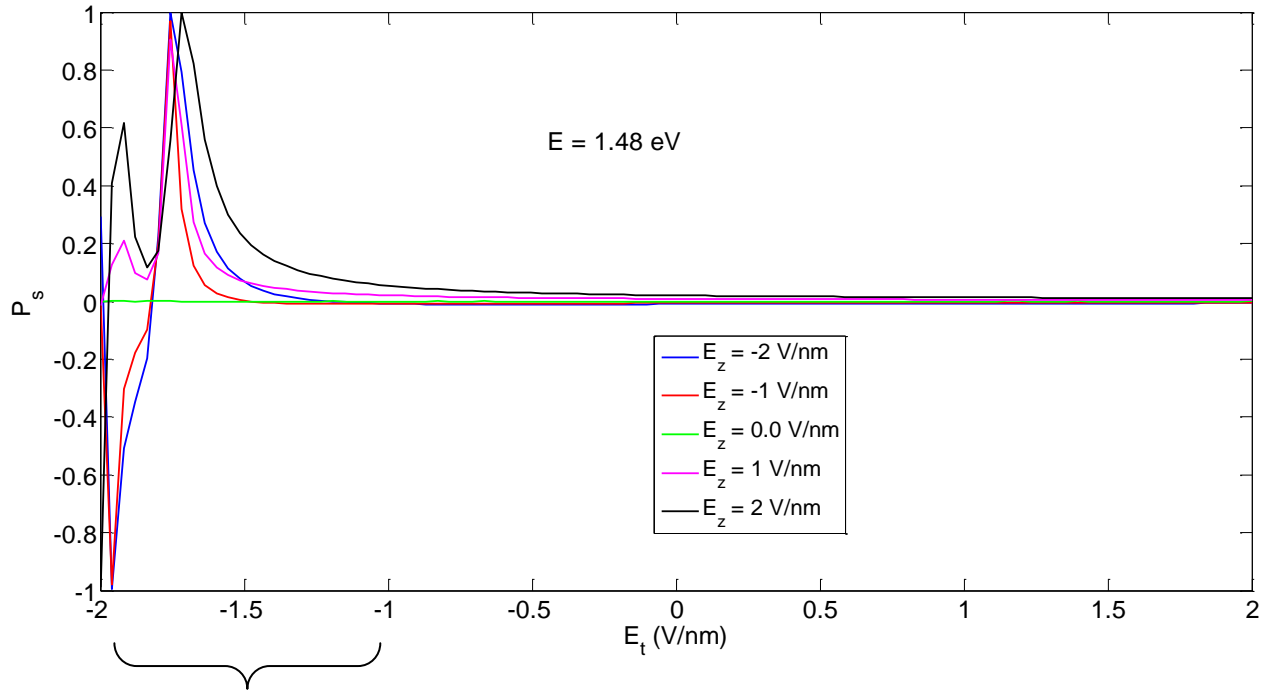

(a)

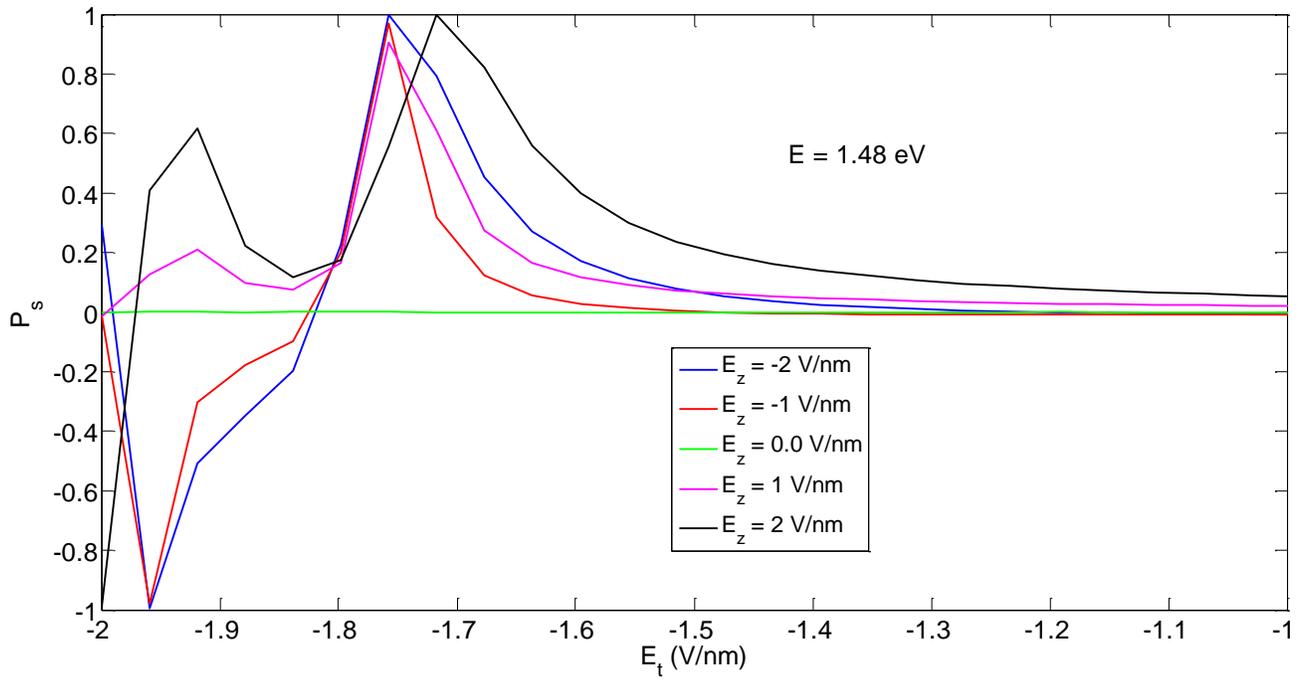

(b)

Fig. 6 (Color online) Spin polarization versus transverse electric field, $E_t$ for different values of perpendicular electric field $E_z = -2, -1, 0.0, 1, 2$ V/nm, when electron energy $E = 1.48$ eV.

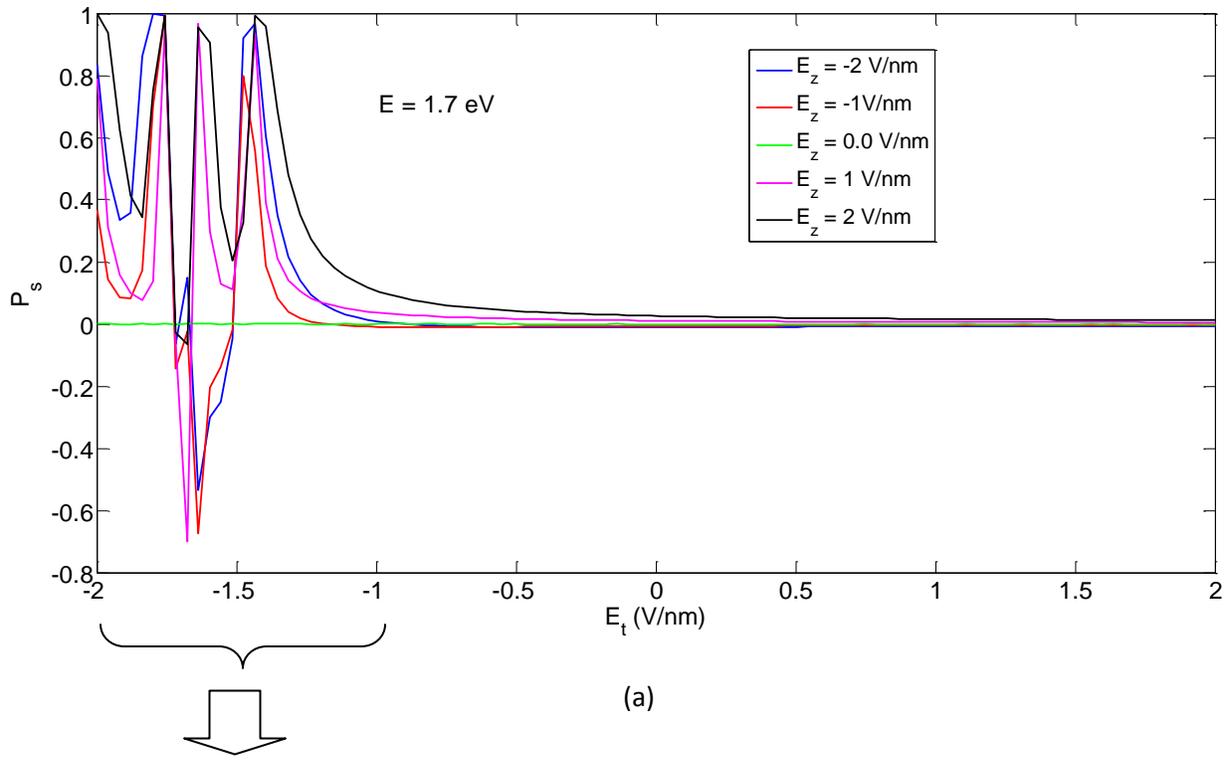

(a)

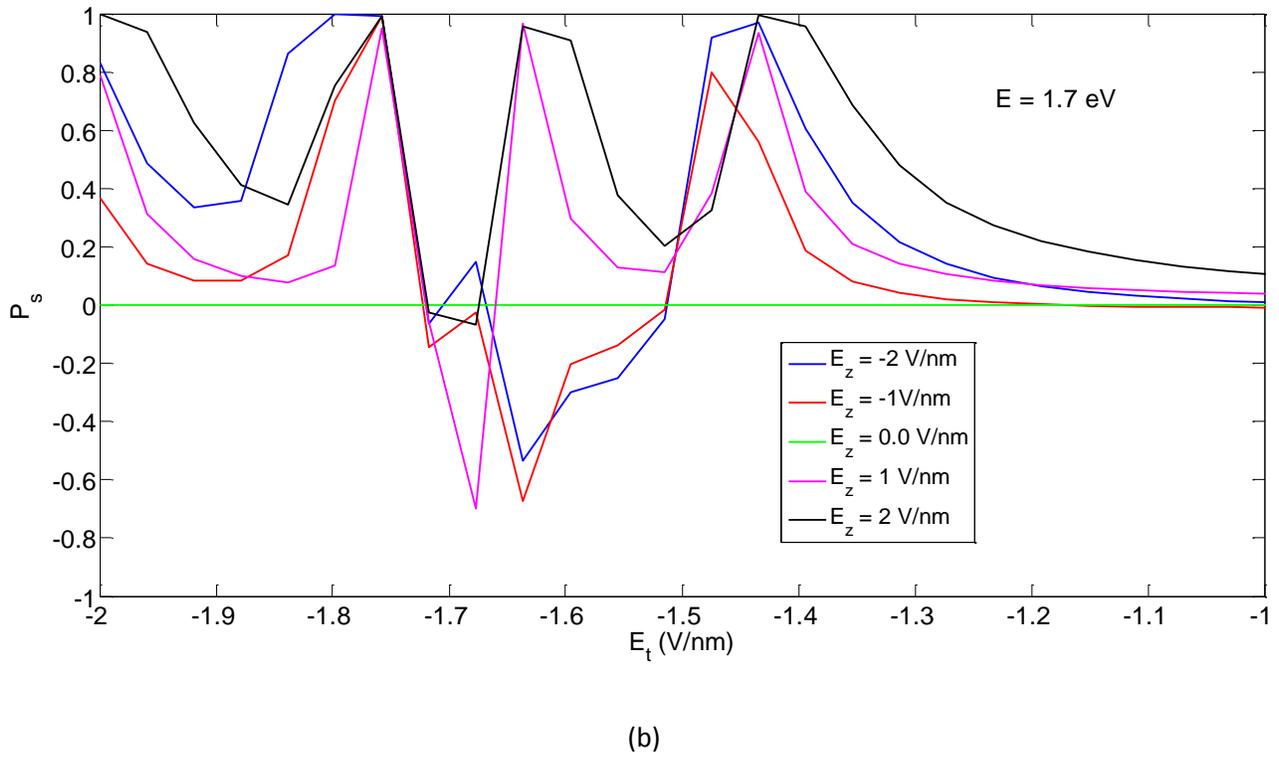

(b)

Fig. 7 (Color online) Spin polarization versus transverse electric field, $E_t$ for different values of perpendicular electric field $E_z = -2, -1, 0.0, 1, 2$ V/nm, when electron energy $E = 1.7$ eV.

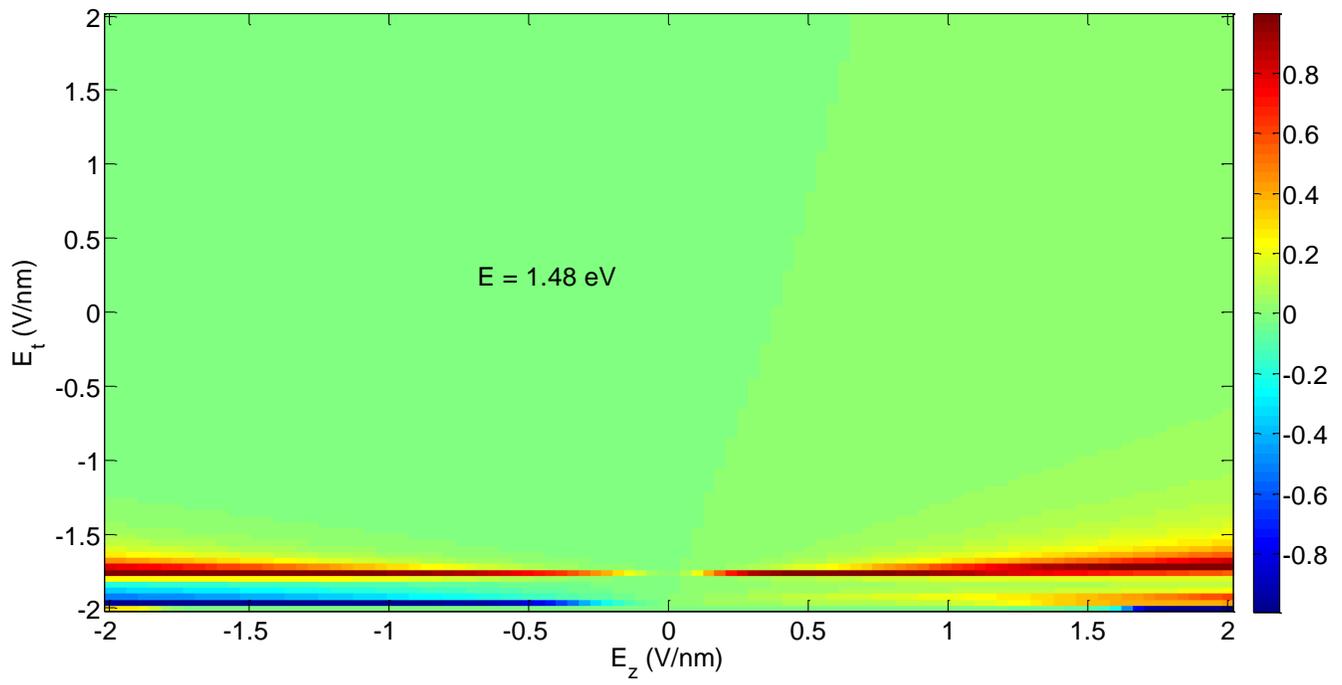

(a)

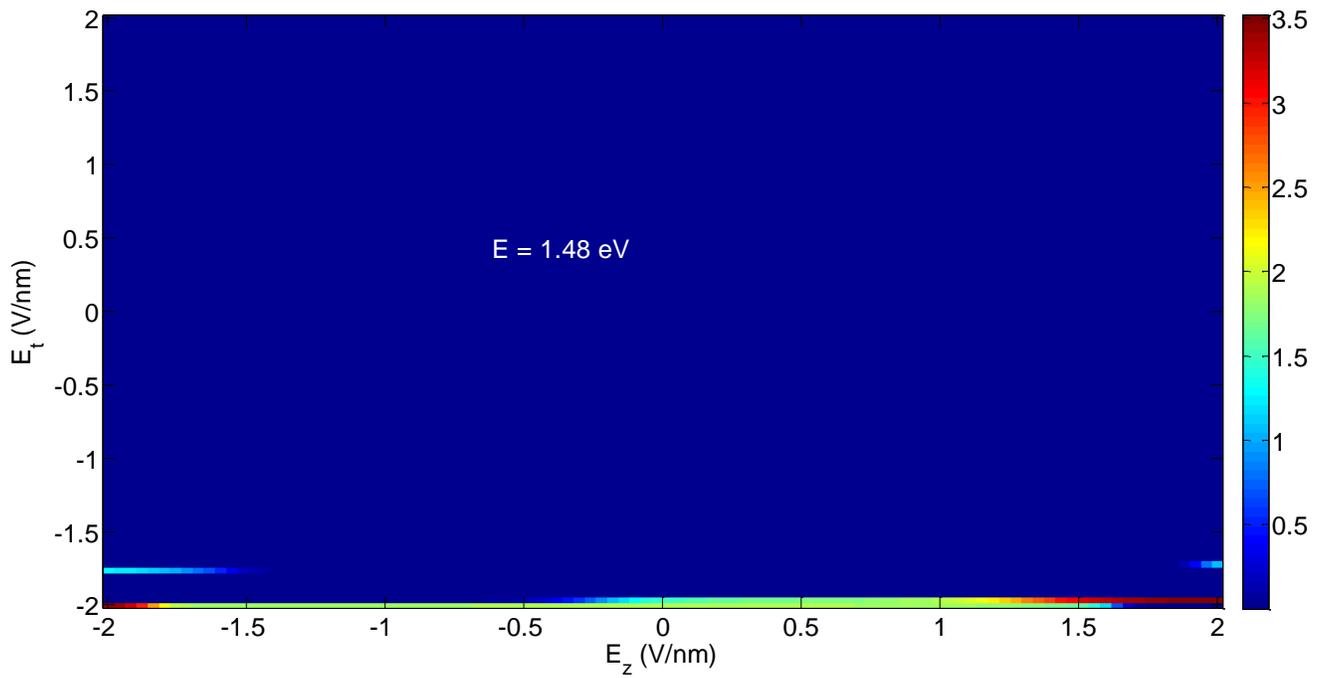

(b)

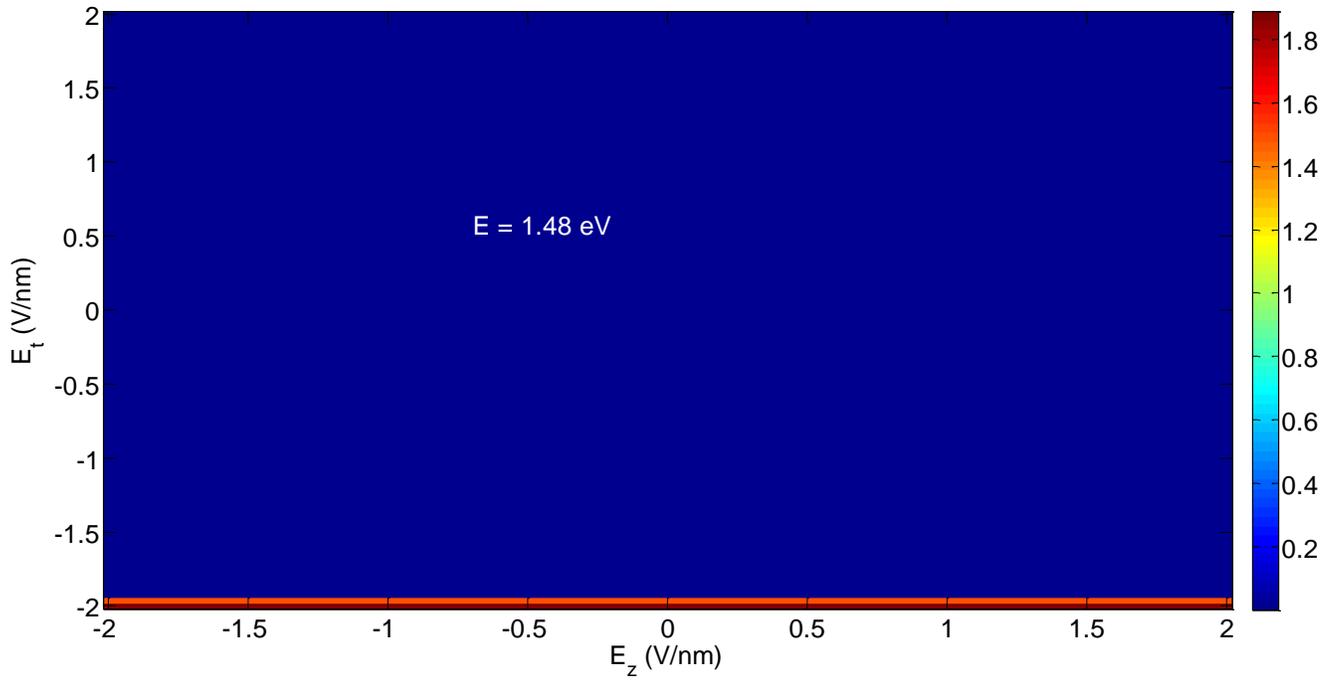

(c)

Fig. 8 (Color online) (a) Spin polarization, (b) conductance of spin up, $G^\uparrow$ and (c) conductance of spin down, $G^\downarrow$ electrons versus transverse electric field, $E_t$ and perpendicular electric field $E_z$ when electron energy $E = 1.48$ eV.

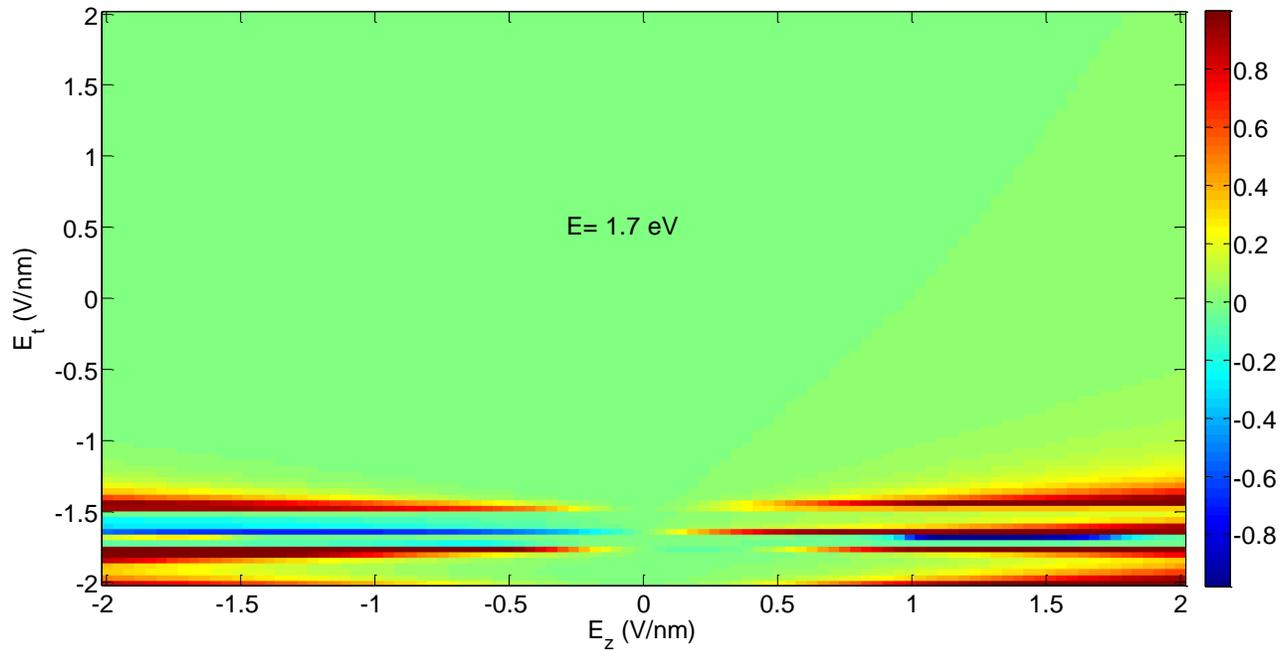

(a)

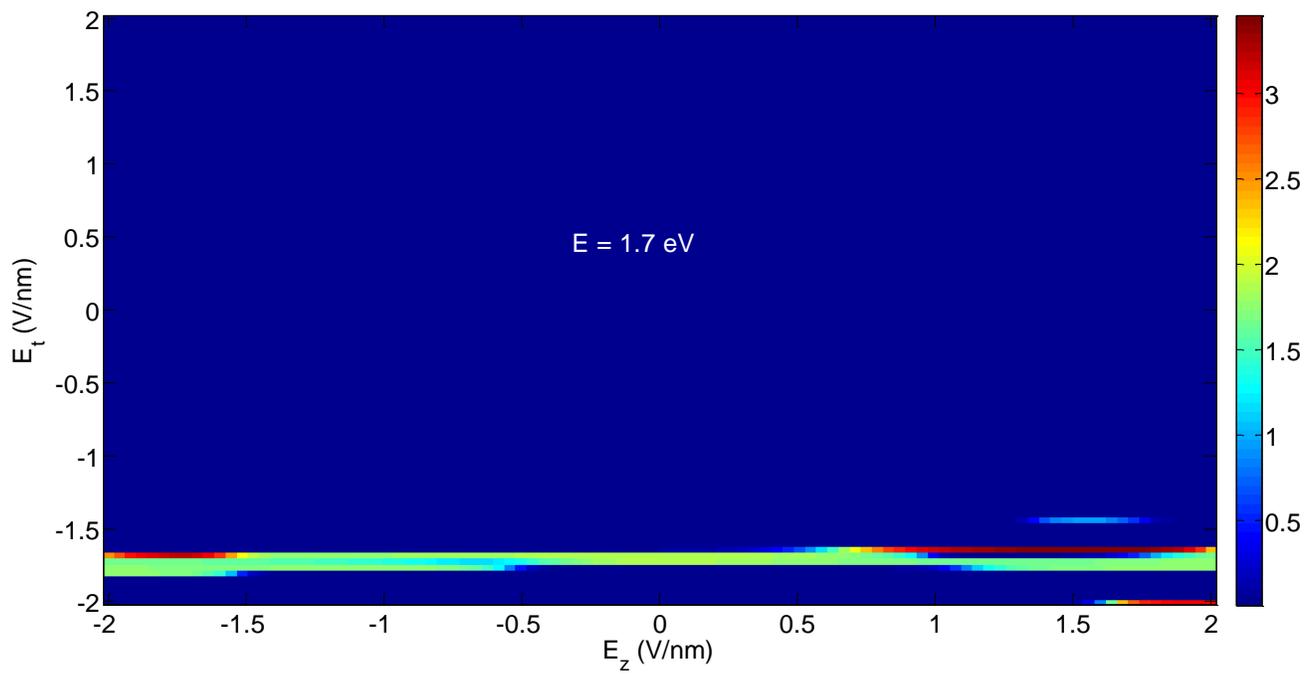

(b)

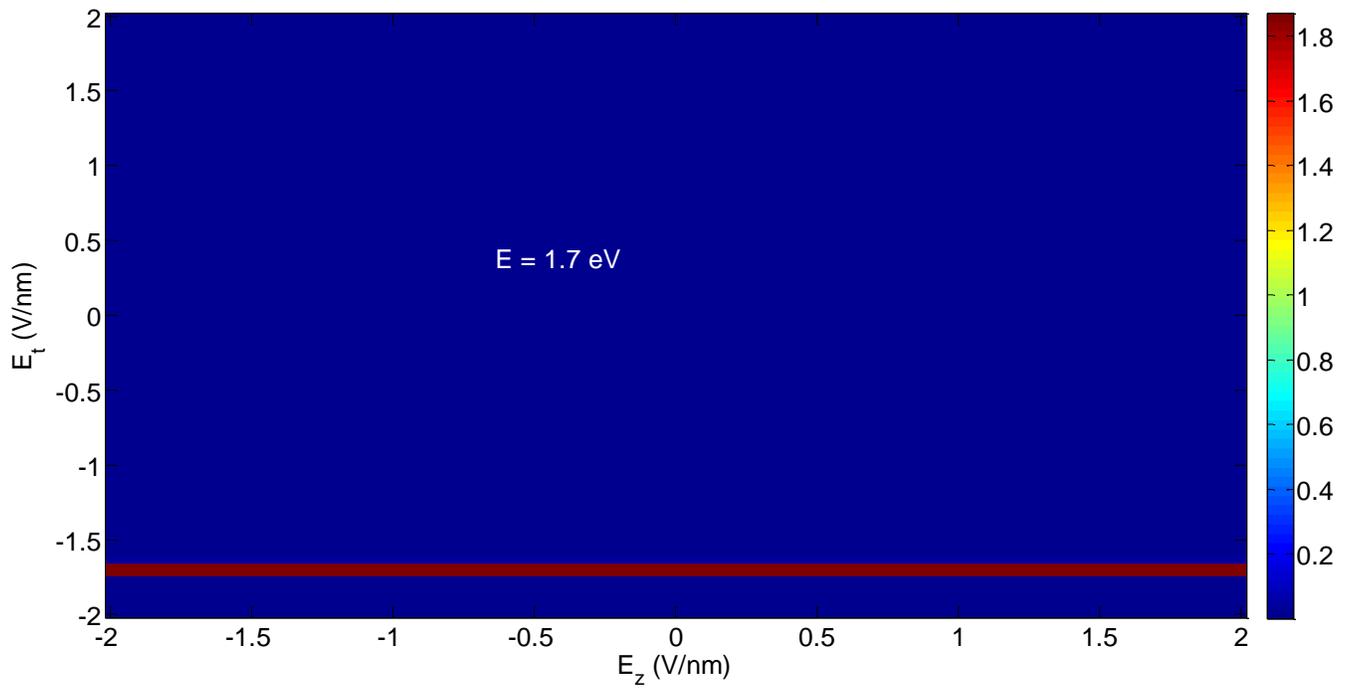

(c)

Fig. 9(Color online) (a) Spin polarization, (b) conductance of spin up, $G^{\uparrow}$ and (c) conductance of spin down, $G^{\downarrow}$ electrons versus transverse electric field, $E_t$ and perpendicular electric field $E_z$ when electron energy $E = 1.7$ eV.